\documentclass[aps,pra,preprint]{revtex4-1}

\usepackage{amsmath}
\usepackage{amssymb}
\usepackage{graphicx}

\begin{document}

\title{Tunable thermoelectricity in monolayers of MoS$_{2}$ and other
group-VI dichalcogenides}

\author{M. Tahir\footnote{m.tahir06@alumni.imperial.ac.uk} and
U. Schwingenschl\"{o}gl\footnote{udo.schwingenschlogl@kaust.edu.sa,+966(0)544700080}}
\affiliation{PSE Division, KAUST, Thuwal 23955-6900, Kingdom of Saudi Arabia}

\begin{abstract}
We study the thermoelectric properties of monolayers of MoS$_{2}$ and other
group-VI dichalcogenides under circularly polarized off-resonant light.
Analytical expressions are derived for the Berry phase mediated magnetic
moment, orbital magnetization, as well as thermal and Nernst conductivities.
Tuning of the band gap by {\it off-resonant} light enhances the spin splitting in
both the valence and conduction bands and, thus, leads to a dramatic
improvement of the spin and valley thermoelectric properties.
\end{abstract}

\maketitle

\affiliation{PSE Division, KAUST, Thuwal 23955-6900, Kingdom of Saudi Arabia}

\section{Introduction}

Being the first truly two dimensional material \cite{1}, graphene has
attracted remarkable attention, both due to its exotic transport behavior
and technological applications in various fields \cite{2}. Still,
fundamental problems restrict its applicability, in particular
the negligible band gap and weak spin orbit coupling (SOC). These limitations
could be overcome by monolayer MoS$_{2}$, which therefore is interesting for
next generation nanoelectronics \cite{3,4,5,6,7}. MoS$_{2}$ combines the
honeycomb structure of graphene with a large intrinsic direct band gap of
$2\Delta=1.66$ eV and a large SOC of $\lambda =74$ meV, providing mass
to the Dirac fermions \cite{8,9,10}. As a consequence, preliminary results
indicate potential in valleytronics, because the dispersion can be
manipulated in a flexible manner for optoelectronic applications \cite%
{10,11,12,13}. Spin and valley Hall effects have been predicted in an
experimentally accessible temperature regime \cite{9}, the former arising
from the strong SOC and the latter from the broken inversion symmetry.

In addition to the electrical and optical properties, Berry phase mediated
thermoelectric effects due to a temperature gradient have been proposed
for two-dimensional systems \cite{15}. Orbital magnetic
moments, orbital magnetizations \cite{17}, as well as thermal and Nernst
conductivities have been addressed in Refs. \cite{19,20,21} and theoretical
models for the thermoelectric transport have been presented for graphene in
Ref. \cite{22} and for topological insulators in Ref.\ \cite{23}. Of
particular interest is the tuning of the spin and valley thermoelectric
properties of MoS$_{2}$ and other group-VI dichalcogenides, where a
temperature gradient gives rise to transverse spin/valley accumulation and
spin/valley current. In graphene this is difficult to realize due to the
negligible band gap and weak SOC.

In the present work we quantify the Berry phase mediated thermoelectric
properties of MoS$_{2}$ and other group-VI dichalcogenides by deriving
analytical expressions for the key thermoelectric quantities in the presence
of circularly polarized {\it off-resonant} light. Gap opening by such light has
been predicted for graphene and for the surface states of topological
insulators \cite{24}, and has been confirmed experimentally for the latter 
\cite{25}. For graphene the chiralities for different values of the frequency have been
given in Ref.\ \cite{new1}. Moreover, opening of a trivial gap has been reported
under high-frequency linearly polarized light \cite{new2}.
Going beyond these findings, we demonstrate in the following that
by off-resonant light large spin and valley thermoelectric effects can be achieved.

\section{Model formulation}

Extending the approach of Ref.\ \cite{9} by introducing time dependence, we
start from the effective Hamiltonian%
\begin{equation}
H^{s_{z},\eta }(t)=v(\eta \sigma _{x}\Pi _{x}(t)+\sigma _{y}\Pi
_{y}(t))+\Delta \sigma _{z}-\lambda \eta s_{z}\sigma _{z}+\lambda \eta s_{z}{\bf I} \label{1}
\end{equation}%
in the $xy$-plane in the presence of circularly polarized light, where $\eta
=\pm 1$ represents the $K$- and $K^{\prime }$-valleys, respectively, $\Delta 
$ is the mass term that breaks the inversion symmetry, $\sigma _{x}$,
$\sigma _{y}$, and $\sigma _{z}$ are the Pauli matrices, $\lambda $ is the
SOC with real spin index $s_{z}$, and $v=at_0/\hbar$ is the Fermi velocity of the Dirac
fermions (with $t_{0}$ being the nearest neighbour hopping amplitude and $a$ the lattice
constant). We use the gauge in the two-dimensional canonical momentum
$\mathbf{\Pi (}t\mathbf{)=p}+e\mathbf{A(}t\mathbf{)}$ with vector potential%
\begin{equation}
\mathbf{A}(t)=(\pm A\sin \Omega t,A\cos \Omega t),  \label{2}
\end{equation}%
where $\Omega $ is the frequency of the light and $A=E_{0}/\Omega $ ($E_{0}$
is the amplitude of the electric field, $\mathbf{E}(t)=-\partial\mathbf{A}%
(t)/\partial t$). We have $\mathbf{A}(t+T_0)=\mathbf{A}(t)$ for $T_0=2\pi
/\Omega $. The plus/minus sign in Eq.\ (2) refers to right/left-handed
circular polarization of the light.

The effect of off-resonant light can be described by a static Floquet
Hamiltonian \cite{24}, which yields excellent agreement with experiments 
\cite{25}. A static approach is satisfied for low intensity ($evA\ll \hslash
\Omega $) and high frequency ($t_{0}\ll \hslash \Omega $) light, which does not
directly excite electrons but effectively modifies the band structure
through virtual photon absorption and emission processes. We arrive at the
effective Hamiltonian (see the Appendix)%
\begin{equation}
H_{eff}^{s_{z},\eta }=v(\eta \sigma _{x}p_{x}+\sigma _{y}p_{y})+(\Delta
+\eta \Delta _{\Omega })\sigma _{z}-\lambda \eta s_{z}\sigma _{z}+\lambda
\eta s_{z}{\bf I},  \label{3}
\end{equation}%
where $\Delta _{\Omega }=e^{2}v^{2}\hslash ^{2}A^{2}/\hslash ^{3}\Omega ^{3}$
is an effective energy term representing the circularly polarized
off-resonant light, which essentially renormalizes the mass of the Dirac
fermions. Similar approaches have been used for describing gapped systems
such as silicene \cite{26} and disordered topological insulators \cite{27}.
Diagonalization of the Hamiltonian leads to the eigenvalues%
\begin{equation}
E_{\zeta }^{s,\eta }=s\eta \lambda +\zeta \sqrt{(v\hslash k)^{2}+(\Delta
+\eta \Delta _{\Omega }-\lambda \eta s)^{2}}  \label{4}
\end{equation}%
and eigenfunctions%
\begin{equation}
\Psi _{\zeta }^{s,\eta }(\mathbf{k})=\frac{e^{ik_{x}x+ik_{y}y}}{\sqrt{%
L_{x}L_{y}}}\left( 
\begin{array}{c}
\frac{v\hslash ke^{-i\eta \varphi }}{\sqrt{(v\hslash k)^{2}+\left[ -\Delta
-\eta \Delta _{\Omega }+\eta s\lambda +\zeta \sqrt{(v\hslash k)^{2}+(\Delta
+\Delta _{\Omega }-\lambda \eta s)^{2}}\right] ^{2}}} \\ 
\frac{-\Delta -\eta \Delta _{\Omega }+\eta s\lambda +\zeta \sqrt{(v\hslash
k)^{2}+(\Delta +\Delta _{\Omega }-\lambda \eta s)^{2}}}{\sqrt{(v\hslash
k)^{2}+\left[ -\Delta -\eta \Delta _{\Omega }+\eta s\lambda +\zeta \sqrt{%
(v\hslash k)^{2}+(\Delta +\Delta _{\Omega }-\lambda \eta s)^{2}}\right] ^{2}}}%
\end{array}%
\right) .  \label{5}
\end{equation}%
Here $\zeta =\pm 1$ denotes electron/hole bands, $s$ = $\pm $1 stands for
spin up/down, and $\varphi =\tan ^{-1}(k_{y}/k_{x})$ with $k_{x}=k\cos
\varphi $, $k_{y}=k\sin \varphi$, and $k=\sqrt{k_{x}^{2}+k_{y}^{2}}$.

The energy eigenvalues $E_{\zeta }^{s,\eta }$ are illustrated in Fig.\ 1. For 
$\hslash \Omega =10t_{0}=11$ eV the circularly polarized light is in the
off-resonance regime, where $\Delta _{\Omega }=0.6$ eV ($evA=2.58$ eV, top
row of Fig. 1) or $\Delta _{\Omega }=0.8$ eV ($evA=2.97$ eV, bottom row of
Fig. 1). The direct band gap of MoS$_{2}$ amounts to $\Delta=0.83$ eV and we have
$\lambda =37$ meV, $v=0.5\times $10$^{5}$ m/s, and $a=3.193$ \AA\ \cite{9}.
For $\Delta _{\Omega }=0.6$ eV the band gap is reduced to 0.46 eV for the $%
K^{\prime }$-valley (top right) and enlarged to 2.86 eV for the $K$-valley
(top left). The effect of the off-resonant light ($\Delta _{\Omega }$) can
be tuned by varying the intensity, where various values have been
achieved experimentally \cite{25}. For $\Delta _{\Omega }=0.8$ eV we observe
that the spin splitting in the conduction band is increased and the band gap
is reduced to 0.06 eV in the $K^{\prime }$-valley, so that only the $%
K^{\prime }$-valley is relevant, whereas for the $K$-valley the band gap
becomes 3.26 eV. In the following we restrict the discussion to the
$K'$-valley ($\eta =-1$).

\section{Orbital magnetic moment and temperature dependent orbital
magnetization}

In order to study the Berry phase mediated thermoelectric transport, we
consider the free energy, which, for a weak magnetic field $\mathbf{B}$, is
given by \cite{15}%
\begin{equation}
F_{\zeta }^{s,-1}=-\frac{1}{\beta }\sum_{\mathbf{k}}\log \left( 1+e^{-\beta
(E_{\zeta }^{s,-1}(\mathbf{k})-\mu )}\right) ,  \label{6}
\end{equation}%
where $\beta =1/k_{B}T$, $k_{B}=8.62\times 10^{-5}$ eV/K is the Boltzmann
constant, $\mu $ is the Fermi energy, and $T$ is the temperature. Eq. (6)
can be simplified by converting the summation into an integral,%
\begin{equation}
F_{\zeta }^{s,-1}=-\frac{1}{\beta }\int \frac{d^{2}k}{(2\pi )^{2}}\left( 1+%
\frac{e}{\hslash }\mathbf{\Omega }_{\zeta }^{s,-1}(\mathbf{k})\cdot \mathbf{B%
}\right) \log \left( 1+e^{-\beta (E_{\zeta }^{s,-1}(\mathbf{k})-\mu
)}\right) ,  \label{7}
\end{equation}
where%
\begin{equation}
\mathbf{\Omega }_{\zeta }^{s,-1}(\mathbf{k})=\mathbf{\nabla }_{\mathbf{k}%
}\times \left\langle \Psi _{\zeta }^{s,-1}(\mathbf{k})\left\vert i\mathbf{%
\nabla }_{\mathbf{k}}\right\vert \Psi _{\zeta }^{s,-1}(\mathbf{k}%
)\right\rangle \mathbf{\hat{z}}  \label{8}
\end{equation}%
is the Berry curvature. The energy $E_{\zeta }^{s,-1}(\mathbf{k})=E_{\zeta
}^{s,-1}-\mathbf{m}_{\zeta }^{s,-1}\mathbf{(k)}\cdot \mathbf{B}$ is modified
by the orbital magnetic moment%
\begin{equation}
\mathbf{m}_{\zeta }^{s,-1}(\mathbf{k})=\frac{-ie}{\hslash }\left\langle
\nabla _{\mathbf{k}}\Psi _{\zeta }^{s,-1}(\mathbf{k})\left\vert \hat{H}%
-E_{\zeta }^{s,-1}\right\vert \nabla _{\mathbf{k}}\Psi _{\zeta }^{s,-1}(%
\mathbf{k})\right\rangle \mathbf{\hat{z}.}  \label{9}
\end{equation}%
The orbital magnetization then is obtained as%
\begin{eqnarray}
\mathbf{M}_{\zeta }^{s,-1} &=&-\left( \frac{\partial F_{\zeta }^{s,-1}}{%
\partial \mathbf{B}}\right) _{\mu ,T}  \label{10} \\
&=&\int \frac{d^{2}k}{(2\pi )^{2}}f(E_{\zeta }^{s,-1}(\mathbf{k}))\mathbf{m}%
_{\zeta }^{s,-1}\mathbf{(k)}+\frac{e}{\beta \hslash }\int \frac{d^{2}k}{%
(2\pi )^{2}}\mathbf{\Omega }_{\zeta }^{s,-1}\mathbf{(k)}\log \left(
1+e^{-\beta (E_{\zeta }^{s,-1}(\mathbf{k})-\mu )}\right) ,  \notag
\end{eqnarray}%
where $f$ is the Fermi distribution function. From Eqs. (4) and (5) we
obtain for the $z$-component of the Berry curvature%
\begin{equation}
\Omega _{\zeta }^{s,-1}(\mathbf{k})=\frac{\hslash ^{2}v^{2}}{2}\frac{\Delta
-\Delta _{\Omega }+s\lambda }{\left[ (v\hslash k)^{2}+(\Delta -\Delta
_{\Omega }+s\lambda )^{2}\right] ^{3/2}}  \label{11}
\end{equation}%
and correspondingly for the $z$-component of the orbital magnetic moment%
\begin{equation}
m_{\zeta }^{s,-1}(\mathbf{k})=\frac{e}{\hslash }E_{\zeta }^{s,-1}\Omega
_{\zeta }^{s,-1}(\mathbf{k}).  \label{12}
\end{equation}%
For finite $\Delta -\Delta _{\Omega }+s\lambda $ the orbital magnetic moment
has a peak at $k=0$. For $\lambda =0$ we obtain for $\Delta -\Delta _{\Omega
}$ = 30 meV for a single valley and $\Delta _{\Omega }=0.8$ eV 
an orbital magnetic moment of 35 Bohr magnetons. High magnetic moments have
been predicted for systems involving orbital degrees of freedom, such as graphene \cite{17}.

Using Eqs.\ (11) and (12) in Eq.\ (10), we obtain for $T\rightarrow 0$ for the
conduction band the $z$-component%
\begin{equation}
M_{+1}^{s,-1}=\frac{e\mu }{2h}\left( 1-\frac{\Delta -\Delta _{\Omega
}+s\lambda }{\mu +s\lambda }\right) ,  \label{13}
\end{equation}%
which again can be enhanced by reducing the band gap by off-resonant light.
Eq.\ (13) reduces to a previous result for gapped graphene in Ref.\ \cite{17}
in the limit of $\lambda =0$ and $\Delta _{\Omega }=0$. As an example, for $%
\mu $ = 0.2 eV we obtain an orbital magnetization (by dividing the results
of Eq.\ (13) by a typical layer thickness of 0.6 nm) of 0.1 Tesla, which is
well detectable in experiments.

In order to evaluate temperature effects, we study the spin and valley
orbital magnetizations $M^{s}=M_{\zeta }^{+1,-1}-M_{\zeta }^{-1,-1}$ and $%
M^{v}=M_{\zeta }^{+1,-1}+M_{\zeta }^{-1,-1}$, using Eq.\ (8), in Fig.\ 2 as a
function of the Fermi energy for temperatures of $T$ = 160 K (left) and $T$
= 360 K (right). To achieve spin and valley orbital magnetizations,
respectively, we require $\lambda >\left\vert \Delta -\Delta _{\Omega
}\right\vert $ and $\lambda <\left\vert \Delta -\Delta _{\Omega }\right\vert 
$. $M^{v}$ is growing faster in the conduction band than in the valence band
and is much/slightly larger than $M^{s}$ in the valence/conduction band.
Increasing the intensity of the off-resonant light increases/decreases $%
M^{v} $ in the conduction/valence band due to a slow/fast reduction of the
band gap, which is due to the energy shift by SOC (see Eq.\ (4)).

As compared to Fig.\ 2, without off-resonant light we obtain about half the
value for $M^{v}$, while $M^{s}$ is 100 times smaller (with opposite sign), since
it is dominated by the band gap (the system is pinned to the valley transport
regime). For $T=160$ K we observe spin effects close to the Dirac point,
whereas they are suppressed for $T=360$ K. These effects will become clear
in the next section, since the orbital magnetization is proportional to the
Hall conductivity \cite{17}. Susceptibility measurements, electron
paramagnetic resonance, x-ray magnetic circular dichroism, and neutron
diffraction can be used to probe the orbital magnetization \cite{30,31,32}.

\section{Spin/valley thermal and Nernst Conductivities}

Eq.\ (10) contains conventional (first term) and Berry phase mediated (second
term) contributions. It has been demonstrated in Refs. \cite{15,17}
that the conventional part does not contribute to the transport, while the
Berry term directly modifies the intrinsic Hall current (which is obtained
by integrating the Berry curvature over the two-dimensional Brillouin zone).
In contrast to the Hall conductivity, the Nernst conductivity is determined
not only by the Berry curvature but also by entropy generation around the
Fermi surface \cite{22,23} and therefore is sensitive to changes of the
Fermi energy and temperature. For a weak electric field $\mathbf{E}$, the
Hall current is given by $j_{x}=\alpha _{xy}^{s,-1}(-\nabla _{y}T)$ and the
Nernst conductivity by \cite{22}%
\begin{equation}
\alpha _{xy}^{s,-1}=\frac{ek_{B}}{\hslash }\sum_{\mathbf{s,\zeta }}\int 
\frac{d^{2}k}{(2\pi )^{2}}\Omega _{\zeta }^{s,-1}(\mathbf{k})S_{\zeta
}^{s,-1}(\mathbf{k}),  \label{14}
\end{equation}%
$S_{\zeta }^{s,-1}(\mathbf{k})=-f(E_{\zeta }^{s,-1}(\mathbf{k})-\mu )\ln
f(E_{\zeta }^{s,-1}(\mathbf{k})-\mu )
-\left( 1-f(E_{\zeta }^{s,-1}(\mathbf{k})-\mu )\right) \ln \left(
1-f(E_{\zeta }^{s,-1}(\mathbf{k})-\mu )\right)$
being the entropy density. Recent experiments found that Eq. (14) describes
graphene very well \cite{33}. We obtain the transverse thermal conductivity
\begin{eqnarray}
\kappa _{xy}^{s,-1} &=&\frac{k_{B}e}{\beta h}\sum_{\mathbf{s,\zeta }}\int 
\frac{d^{2}k}{(2\pi )^{2}}\Omega _{\zeta }^{s,-1}(\mathbf{k})\left\{ \frac{%
\pi ^{2}}{3}+\beta ^{2}(E_{\zeta }^{s,-1}(\mathbf{k})-\mu )^{2}f(E_{\zeta
}^{s,-1}(\mathbf{k})-\mu )\right.   \label{15} \\
&&\left. -\ln ^{2}\left( 1+e^{-\beta (E_{\zeta }^{s,-1}(\mathbf{k})-\mu
)}\right) -2\text{Li}_{2}\left( 1-f(E_{\zeta }^{s,-1}(\mathbf{k})-\mu
)\right) \right\} ,  \notag
\end{eqnarray}%
where Li$_{2}$ is the polylogarithm. Eqs. (14) and (15) can be simplified in
the limit of low temperature, using Mott relations \cite{15}, to%
\begin{equation}
\alpha _{xy}^{s,-1}=-\frac{\pi ^{2}}{3}\frac{k_{B}^{2}}{e}T\frac{d\sigma
_{xy}^{s,-1}}{d\mu }=-e\frac{d\kappa _{xy}^{s,-1}}{d\mu }  \label{16}
\end{equation}%
and%
\begin{equation}
\kappa _{xy}^{s,-1}=\frac{\pi ^{2}}{3}\frac{k_{B}^{2}}{e^{2}}T\sigma
_{xy}^{s,-1},  \label{17}
\end{equation}%
where%
\begin{equation}
\sigma _{xy}^{s,-1}=\frac{e^{2}}{\hslash }\int \frac{d^{2}k}{(2\pi )^{2}}%
\left( f_{+}(E)-f_{-}(E)\right) \Omega _{\zeta }^{s,-1}(\mathbf{k})
\label{18}
\end{equation}%
with $f_{\pm }$ representing the distribution function of the electron/hole
bands. According to Eq. (16), the Nernst conductivity is proportional to the
derivative of the thermal conductivity. We solve Eq.\ (18) for $T=0$
by performing the integral and obtain in the case that the Fermi
level is in the conduction band%
\begin{equation}
\sigma _{xy}^{s,-1}=-\frac{e^{2}}{2h}\frac{\Delta -\Delta _{\Omega
}+s\lambda }{\mu +s\lambda }.  \label{19}
\end{equation}%
Eqs.\ (14) and (19) show that the spin ($\alpha _{xy}^{s}=\alpha
_{xy}^{+1,-1}-\alpha _{xy}^{-1,-1}$) and valley ($\alpha _{xy}^{v}=\alpha
_{xy}^{+1,-1}+\alpha _{xy}^{-1,-1}$) Nernst conductivities are enhanced
under off-resonant light. Without off-resonant light the spin Nernst
conductivity is negligible because of the vanishing spin Hall conductivity
in the limit $\Delta \gg \lambda $ (the system is pinned to the valley Hall
regime). Due to the large band gap, the valley Nernst conductivity is also
small for $t_{0}<\Delta $. An enhanced spin splitting and corresponding
giant thermoelectric transport in both the conduction and valence bands is
achieved by reducing the band gap (to the range of $\lambda $). The results
in Eqs.\ (14) and (19) guarantee for monolayers of MoS$_{2}$ and related
group-VI dichalcogenides an electrically tunable band gap to tailor the spin
and valley transport.

Experiments indicate that the thermoelectric properties can be understood by
Mott relations, which agree with
experimental data for low temperature \cite{33,34,35,36}. In these experiments the
thermoelectric properties, in particular the Nernst conductivity, have been
measured for gapless graphene in a transverse magnetic field. The Nernst
effect discussed in our work exists even without external magnetic field,
being solely driven by the effective magnetic field due to the Berry
curvature. Note that Eq.\ (14) is more general than Eq.\ (16), because it goes
beyond the linear temperature dependence.

It has been found experimentally that the dependence of the thermoelectric
transport on the gate voltage (Fermi energy) can be tuned by controlling the
band gap in monolayer \cite{19,20,21,33,34,35} and bilayer \cite{36}
graphene. Being the electrical response to the thermodynamic perturbation,
a giant thermoelectric transport is achieved when the bands come close to
the Dirac point. In Fig.\ 3 (top) we show numerical results for
$\alpha _{xy}^{s}$, by evaluating Eq.\ (14), as function of the Fermi
energy at $T=160$ K (left) and $T=360$ K (right) and vary the band gap
by off-resonant light as $\Delta _{\Omega }=0.8$ eV (blue), $\Delta _{\Omega
}=0.81$ eV (green), and $\Delta _{\Omega }=0.82$ eV (red). For $T=160$ K
we find two peaks with negative values, where the lower peak is a consequence
of the spin splitting and is washed out for $T=360$ K.
Our results show that a significant spin dependent
transport can be observed in MoS$_{2}$ and related group-VI
dichalcogenides at room temperature (or even above). Figure 3 (bottom)
shows $\alpha_{xy}^{v}$ as a function
of the Fermi energy for $T=160$ K (left) and $T=360$ K (right) with $\Delta _{\Omega }=0.6$
eV (blue), $\Delta _{\Omega }=0.65$ eV (green), and $\Delta _{\Omega }=0.7$ eV (red).
We observe shifts of the peaks towards the Dirac point for increasing $\Delta _{\Omega }$,
which reflects the reduction of the band gap. The amplitude
grows for decreasing band gap. For $\Delta _{\Omega }=0.8$ to 0.82 eV $\alpha _{xy}^{s}$
is enhanced, whereas below $\Delta _{\Omega }=0.793$ eV we are in the valley
transport regime and obtain an enhancement of $\alpha _{xy}^{v}$.
The transport is huge as compared to the case without off-resonant light.

In general, it depends on the sign of the Berry curvature (compare Fig.\ 1)
whether the Nernst conductivity is positive or negative. Our results are valid
for elevated temperature in the experimentally relevant range 
\cite{33}. Furthermore, $\alpha _{xy}^{s,-1}$ $\neq 0$ when the Fermi energy
is in the band gap, whereas Eq. (16) yields $\alpha _{xy}^{s,-1}=0$ (being
the derivative of $\sigma _{xy}^{s,-1}$, which is quantized and independent
of $\mu $ in this case). The demonstrated enhancement of the spin/valley
transport due to the tunability of the band gap in MoS$_{2}$ and other
group-VI dichalcogenides has been desired for thermoelectric applications
since the discovery of graphene. Band gap opening by off-resonant light has
been achieved in Ref.\ \cite{25} for the surface states of topological
insulators and can also be used for monolayer MoS$_{2}$, since transistors 
\cite{5} and amplifiers \cite{6} already have been realized.

\section{Conclusion}

We have derived analytical results for the thermoelectric transport in
monolayer MoS$_{2}$ and related group-VI dichalcogenides in the presence of
off-resonant light. We have shown that an increased intensity of the light
reduces the direct band gap and results in a strong spin splitting in the
conduction band and, therefore, in dramatic enhancement of the thermoelectric
transport. The spin splitting in the conduction band (in contrast to the
valence band) is negligible without external perturbation (such
as off-resonant light). The band gap even can be tuned to zero with giant
spin splitting in both the valence and conduction bands. The enhancement of
the spin/valley transport properties demonstrated here is desired for
spin/valley dependent thermoelectric devices, whereas the tunable band gap opens new
directions for fundamental transport experiments.

\appendix
\numberwithin{equation}{section}
\section{}
The time dependence in Eq. (1) can be understood as the sum of two
second-order virtual photon processes, where electrons absorb and then emit
a photon and electrons emit and then absorb a photon \cite{24}. Within
Floquet theory, we use the fact that%
\begin{equation}
H_{eff}^{s_{z},\eta }=\frac{i}{T_0}\log [U(T_0)]  \label{A.1}
\end{equation}%
with%
\begin{equation}
U(T_0)=\text{\^{T}}\exp \left[ -i\int_{0}^{T_0}H^{s_{z},\eta }(t)dt\right]
\simeq \exp [-iH_{eff}^{s_{z},\eta }T_0],  \label{A.2}
\end{equation}%
where \^{T} is the time ordering operator. We consider the Fourier
decomposition%
\begin{equation}
H^{s_{z},\eta }(t)=\sum_{n=-\infty }^{\infty }H_{n}^{s_{z},\eta }e^{in\Omega
t}\simeq H_{0}^{s_{z},\eta }+H_{1}^{s_{z},\eta }e^{i\Omega
t}+H_{-1}^{s_{z},\eta }e^{-i\Omega t},  \label{A.3}
\end{equation}%
where we have used%
\begin{equation}
H_n^{s_{z},\eta }=\frac{1}{T_0}\int_{0}^{T_0}e^{-{\rm sgn}(n) i\Omega t}H^{s_{z},\eta
}(t)dt.  \label{A.4}
\end{equation}%
Expanding the exponential function in Eq. (A.2) in a Taylor series yields%
\begin{equation}
\exp \left[ -i\int_{0}^{T_0}H^{s_{z},\eta }(t)dt\right] =1-i%
\int_{0}^{T_0}H^{s_{z},\eta }(t)dt+\frac{(-i)^{2}}{2}\int_{0}^{T_0}H^{s_{z},\eta
}(t_{1})dt_{1}\int_{0}^{T_0}H^{s_{z},\eta }(t_{2})dt_{2}+...  \label{A.5}
\end{equation}%
and thus%
\begin{equation}
U(T_0)\simeq \text{\^{T}}\left( 1-i\int_{0}^{T_0}H^{s_{z},\eta }(t)dt+\frac{%
(-i)^{2}}{2}\int_{0}^{T_0}H^{s_{z},\eta
}(t_{1})dt_{1}\int_{0}^{T_0}H^{s_{z},\eta }(t_{2})dt_{2}\right).  \label{A.6}
\end{equation}%
By applying the time ordering operator in Eq. (A.6) we arrive at\bigskip 
\begin{eqnarray}
U(T_0) &\simeq &1-i\int_{0}^{T_0}H^{s_{z},\eta }(t)dt-\frac{1}{2}%
\int_{0}^{T_0}dt_{1}\int_{0}^{T_{1}}dt_{2}H^{s_{z},\eta }(t_{1})H^{s_{z},\eta
}(t_{2})  \label{A.7} \\
&&-\frac{1}{2}\int_{0}^{T_0}dt_{2}\int_{0}^{T_{2}}dt_{1}H^{s_{z},\eta
}(t_{2})H^{s_{z},\eta }(t_{1}).  \notag
\end{eqnarray}%
Executing the integration in Eq.\ (A.7) with the help of Eq.\ (A.3) and
reordering the terms, we obtain%
\begin{align}
U(T_0)& \simeq 1-iH_{0}^{s_{z},\eta }T_0-\frac{T_0}{\Omega }\left( \pi
(H_{0}^{s_{z},\eta })^{2}-i([H_{0}^{s_{z},\eta },H_{-1}^{s_{z},\eta
}]-[H_{0}^{s_{z},\eta },H_{1}^{s_{z},\eta }]+[H_{-1}^{s_{z},\eta
},H_{1}^{s_{z},\eta }])\right)  \label{A.8} \\
& =1-iH_{eff}^{s_{z},\eta }T_0-\frac{1}{2}(H_{eff}^{s_{z},\eta
}{})^{2}T_0^{2}+...  \notag
\end{align}%
and thus%
\begin{equation}
H_{eff}^{s_{z},\eta }=H_{0}^{s_{z},\eta }+\frac{1}{\hslash \Omega }\left(
[H_{0}^{s_{z},\eta },H_{+1}^{s_{z},\eta }]-[H_{0}^{s_{z},\eta
},H_{-1}^{s_{z},\eta }]\right) +\frac{1}{\hslash \Omega }[H_{+1}^{s_{z},\eta
},H_{-1}^{s_{z},\eta }]  \label{A.9}
\end{equation}%
with $H_{0}^{s_{z},\eta }=$ $v(\eta \sigma _{x}p_{x}+\sigma
_{y}p_{y})+\Delta \sigma _{z}-\lambda \eta s_{z}\sigma _{z}+\lambda \eta
s_{z}$, which describes a static honeycomb lattice with hopping $t_{0}$ (in
a standard tight binding notation) and a band gap of 2$\Delta $. When $%
\hslash \Omega \gg t_{0}$ and $\Delta $ then the off-resonance condition is
satisfied and perturbation theory can be applied. Since the first order term
in Eq. (A.9) vanishes, we simplify the second order term to arrive at the
effective Hamiltonian of Eq.\ (3).

\begin{acknowledgments}
Research reported in this publication was supported by the King Abdullah
University of Science and Technology (KAUST).
\end{acknowledgments}

\begin{figure}[ht]
\begin{center}
\includegraphics[width=0.45\columnwidth,clip]{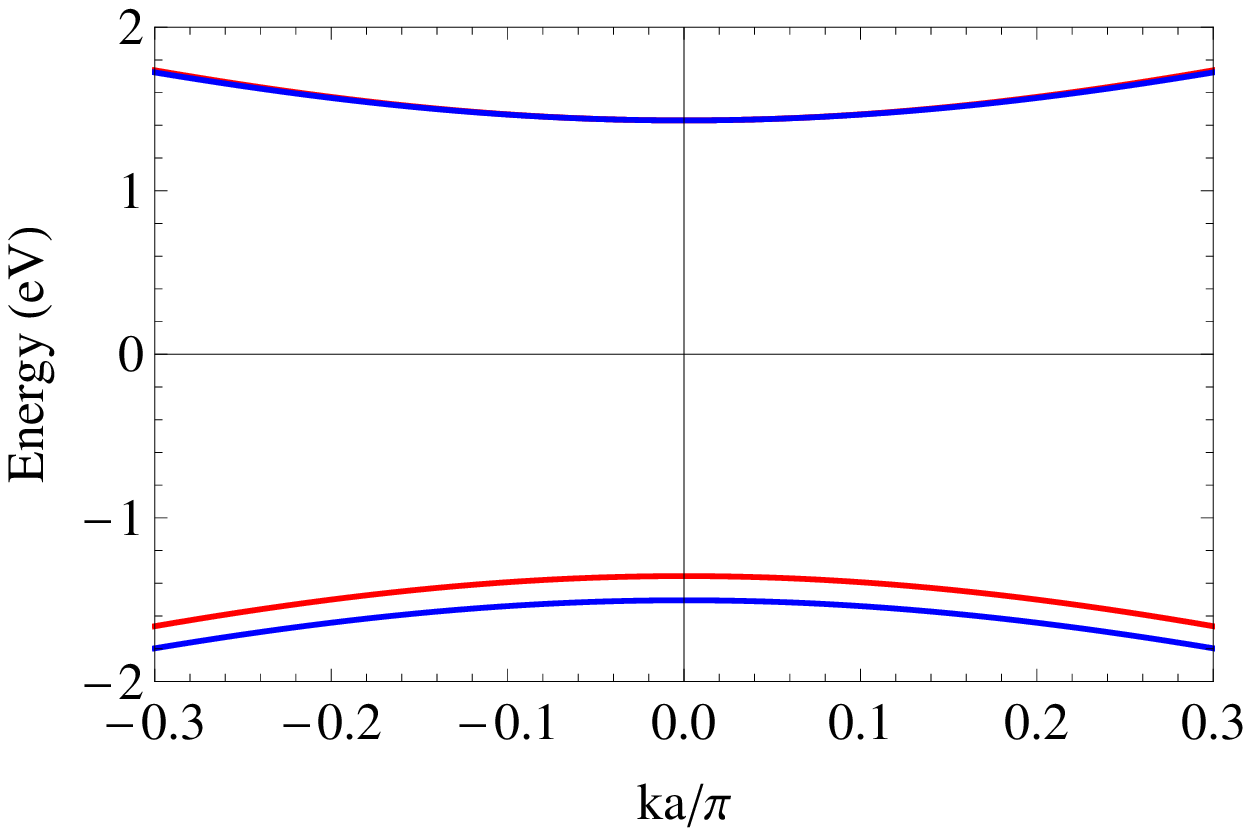} %
\includegraphics[width=0.45\columnwidth,clip]{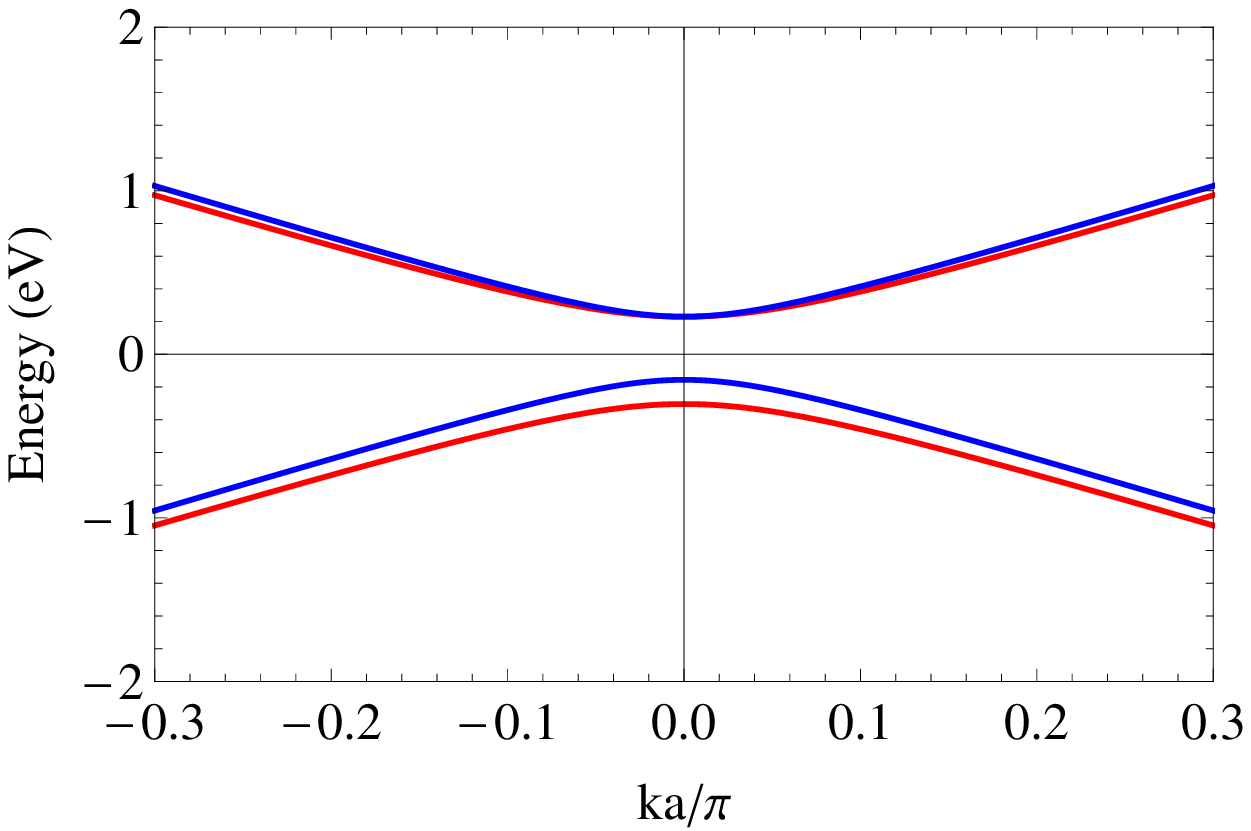} %
\includegraphics[width=0.45\columnwidth,clip]{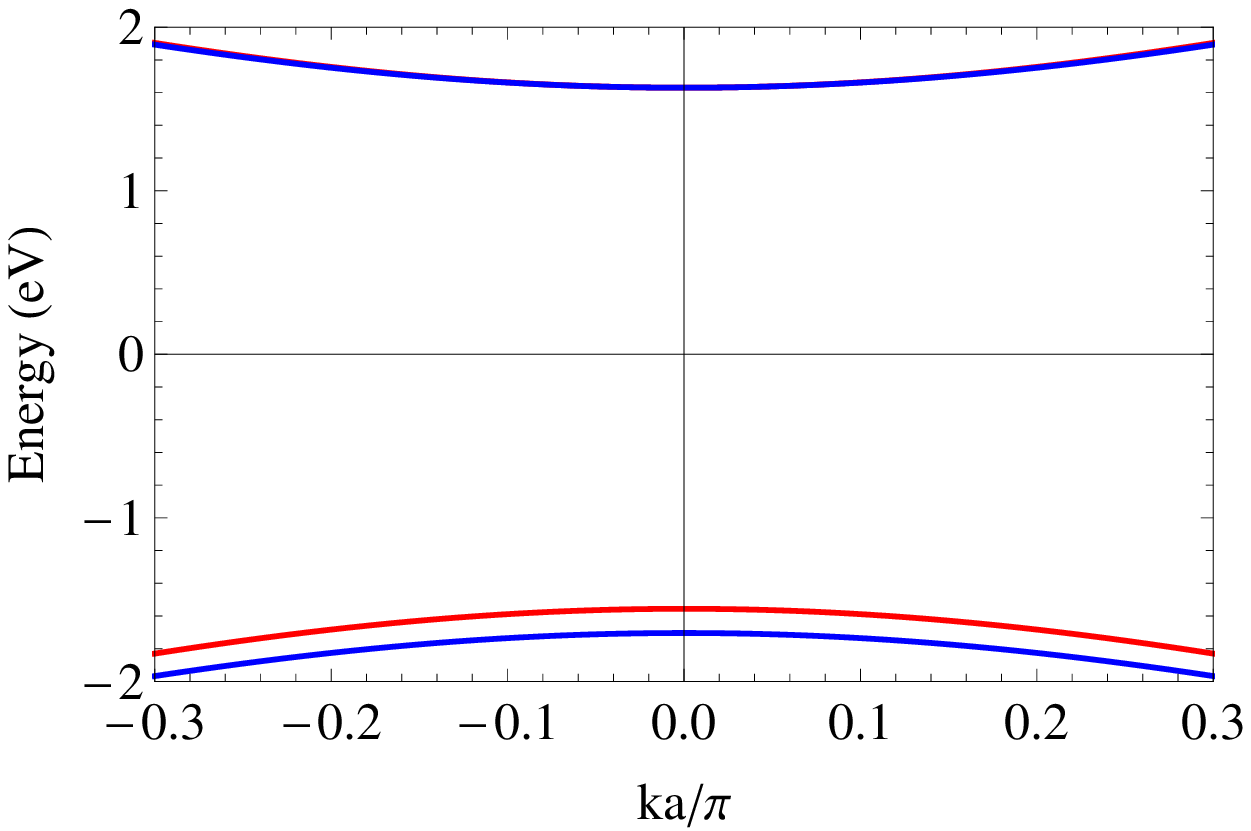} %
\includegraphics[width=0.45\columnwidth,clip]{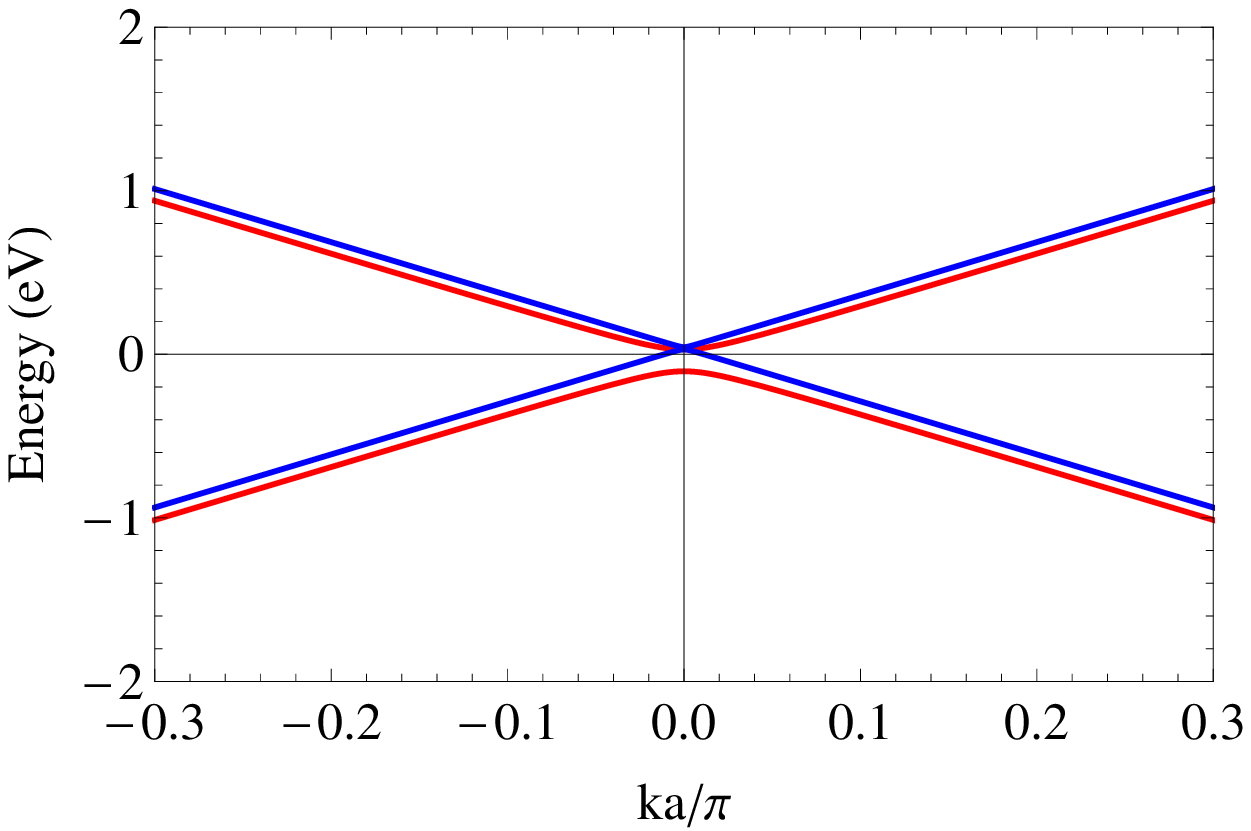}
\end{center}
\caption{Band structure of monolayer MoS$_{2}$ in the presence of
off-resonant light and SOC for the $K$-valley (left) and $K^{\prime }$%
-valley (right) using $\hslash \Omega =10t_{0}$ = 11 eV, $\Delta $ = 0.83
eV, $\protect\lambda =37$ meV, $v=0.5\times $10$^{5}$ m/s, and $a=3.193$ \AA %
. The top row refers to $\Delta _{\Omega }=0.6$ eV and the bottom row to $%
\Delta _{\Omega }=0.8$ eV.}
\label{fig:1}
\end{figure}

\begin{figure}[ht]
\begin{center}
\includegraphics[width=0.45\columnwidth,clip]{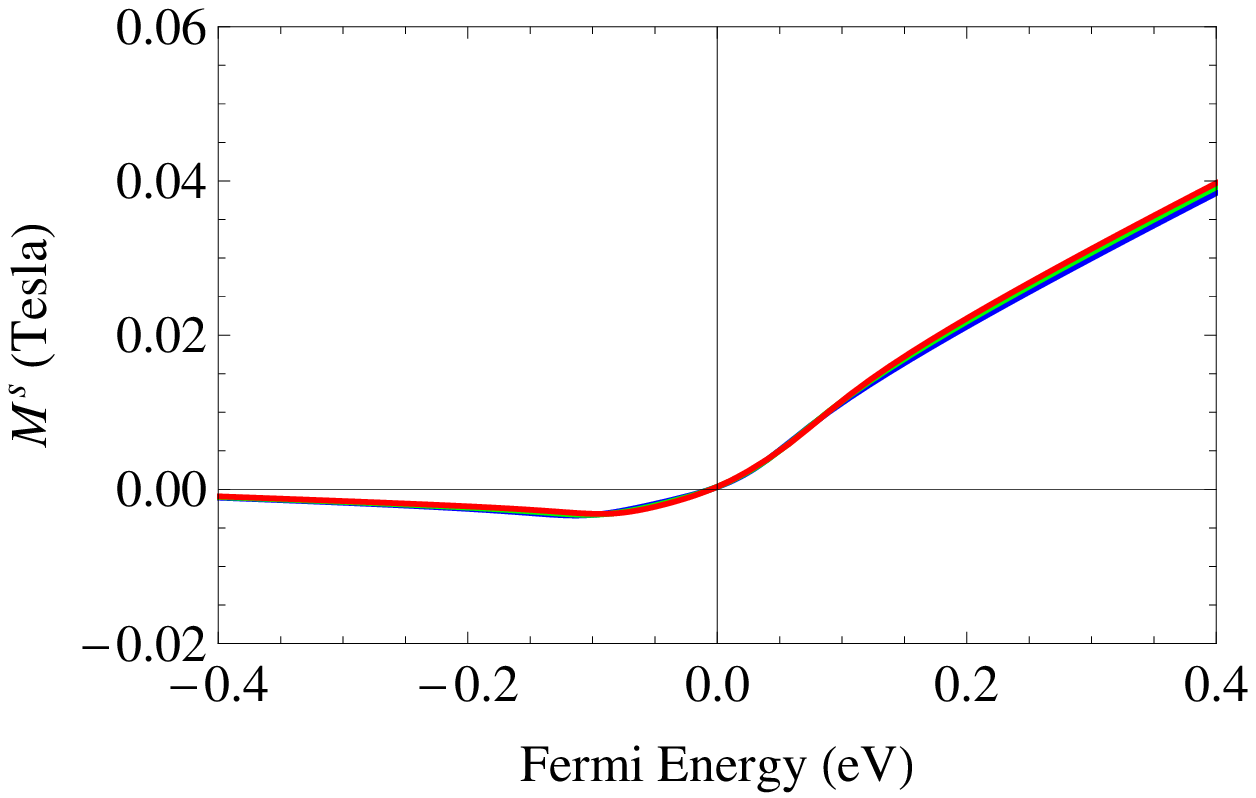} %
\includegraphics[width=0.45\columnwidth,clip]{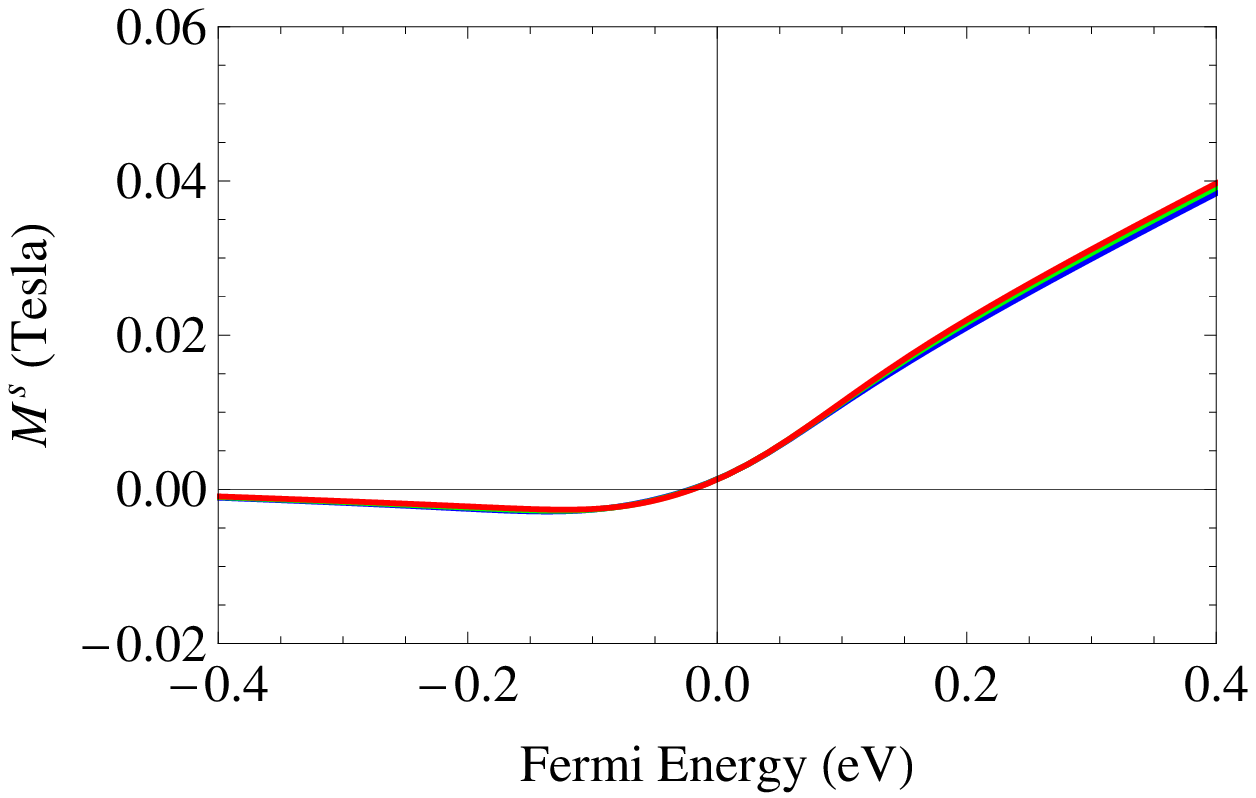} %
\includegraphics[width=0.45\columnwidth,clip]{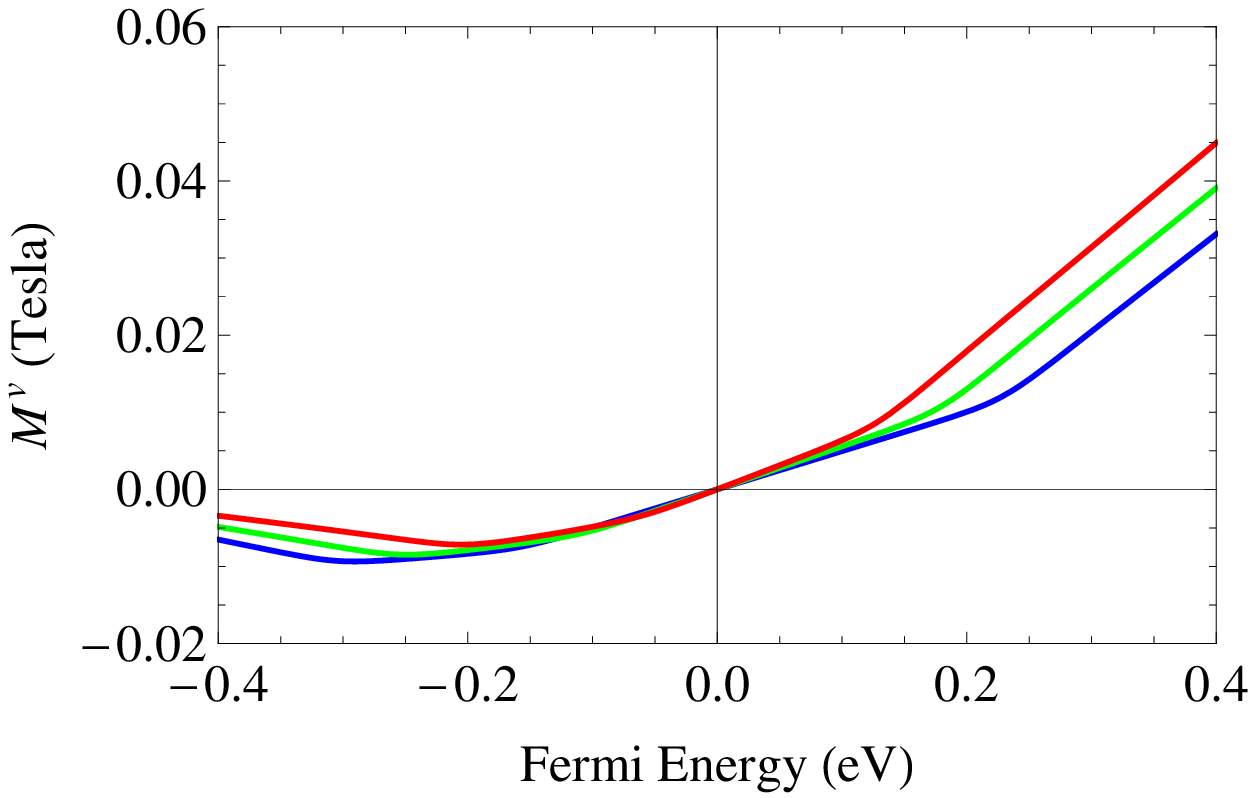} %
\includegraphics[width=0.45\columnwidth,clip]{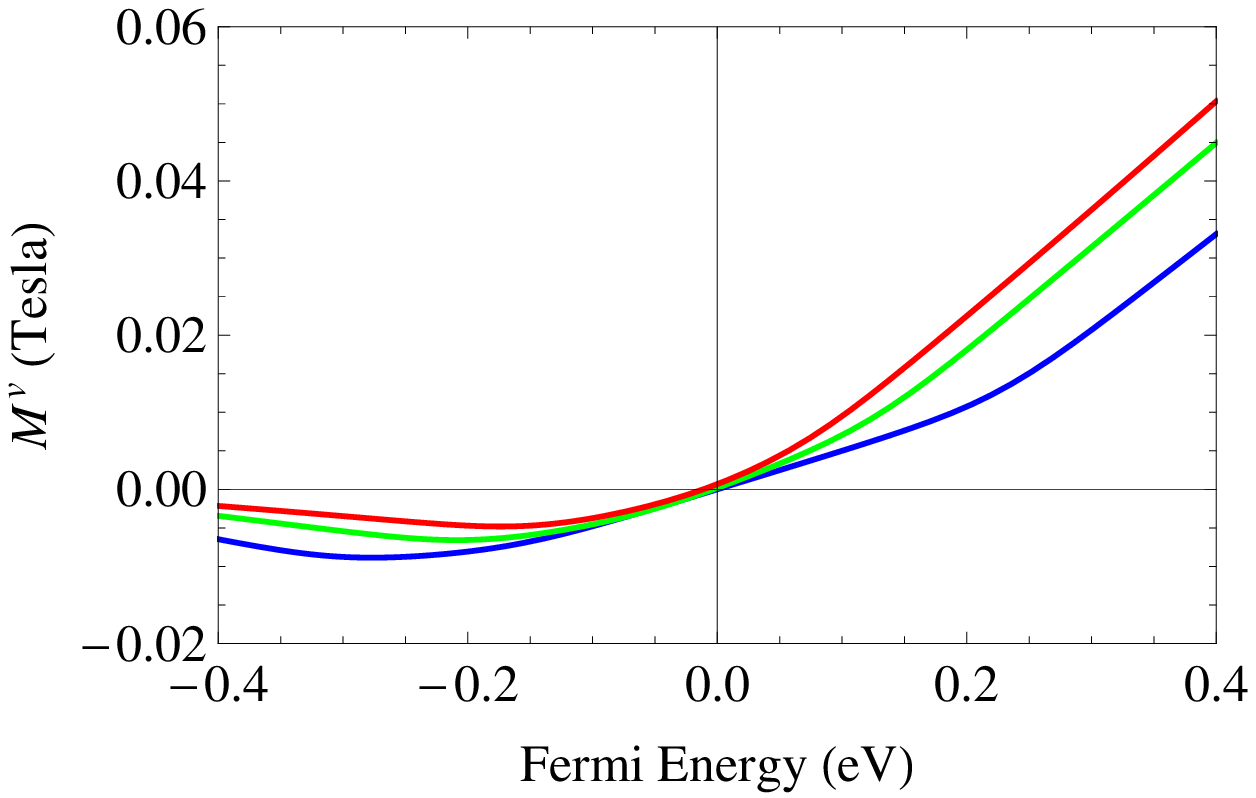}
\end{center}
\caption{Spin (top) and valley (bottom) orbital magnetizations (divided by
the layer thickness 0.6 nm) as a function of the Fermi energy for $T$ =
160 K (left) and $T$ = 360 K (right). For $M^{s}$ we use $\Delta _{\Omega
}=0.8$ eV (blue), $\Delta _{\Omega }=0.81$ eV (green), and $\Delta _{\Omega
}=0.82$ eV (red), whereas for $M^{v}$ we use $\Delta _{\Omega }=0.6$ eV
(blue), $\Delta _{\Omega }=0.65$ eV (green), and $\Delta _{\Omega }=0.7$ eV
(red). All other parameters are the same as in Fig.\ 1.}
\label{fig:2}
\end{figure}

\begin{figure}[ht]
\begin{center}
\includegraphics[width=0.45\columnwidth,clip]{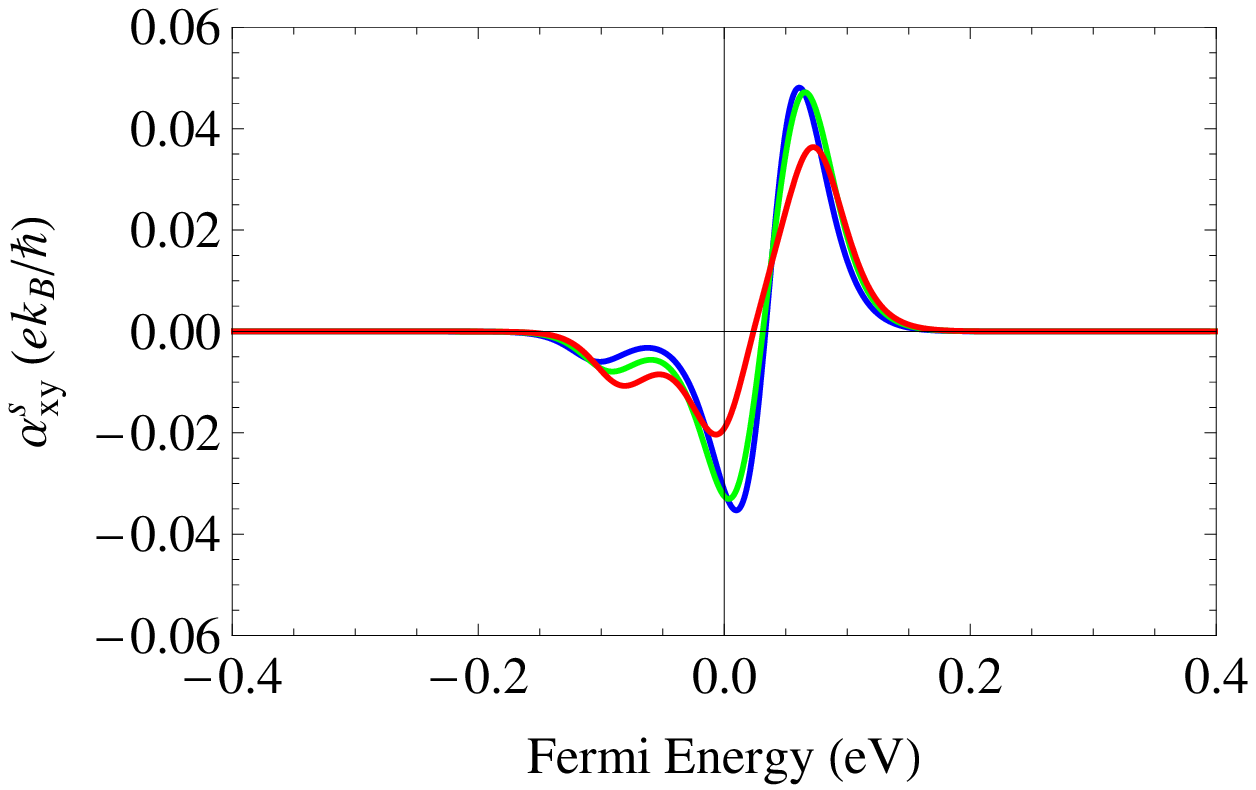} %
\includegraphics[width=0.45\columnwidth,clip]{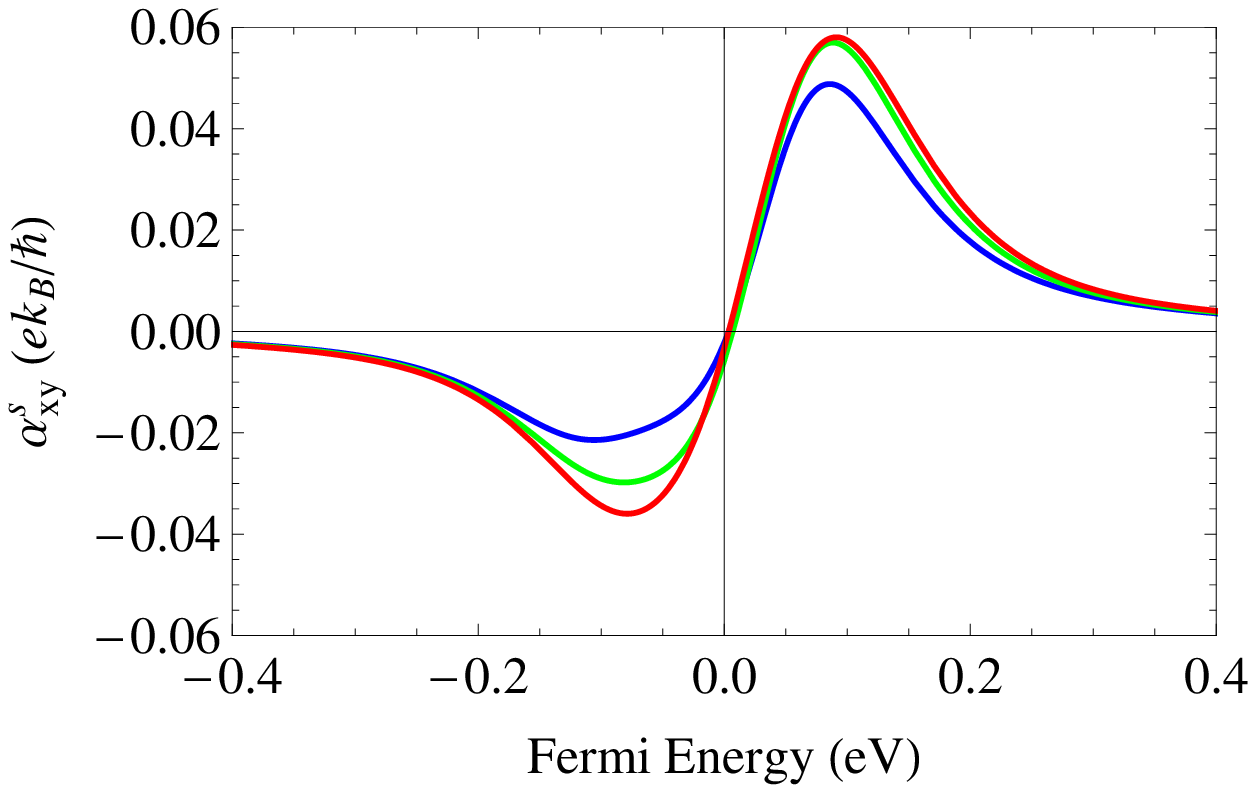} %
\includegraphics[width=0.45\columnwidth,clip]{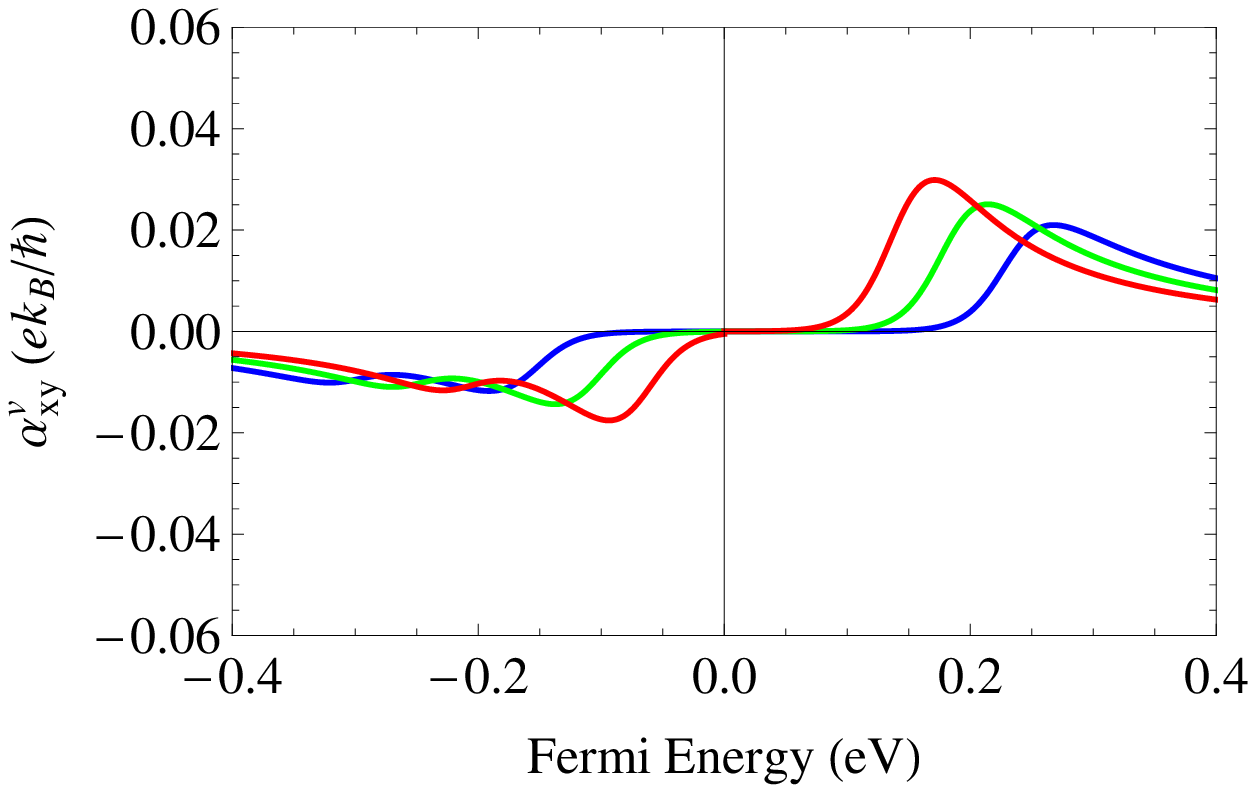} %
\includegraphics[width=0.45\columnwidth,clip]{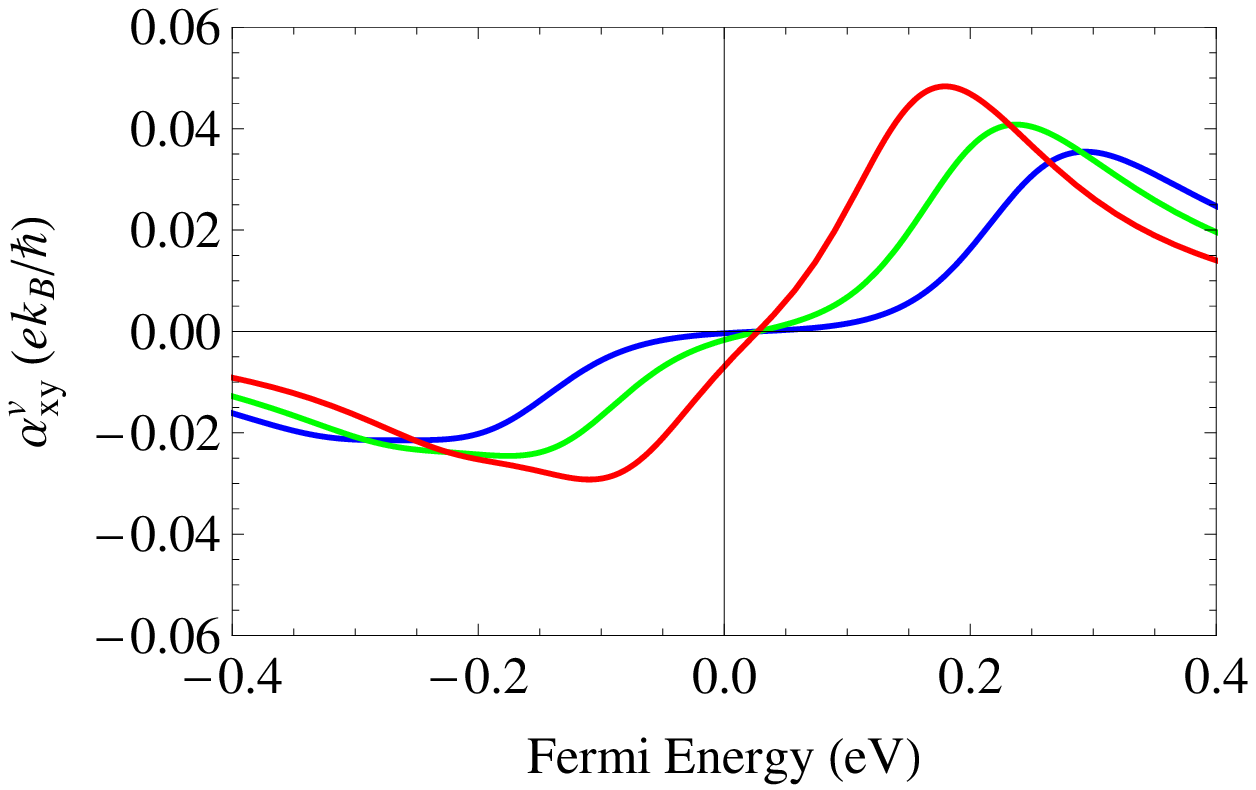}
\end{center}
\caption{Spin (top) and valley (bottom) Nernst conductivities as a function
of the Fermi energy for $T$ = 160 K (left) and $T$ = 360 K (right). For $%
\protect\alpha _{xy}^{s}$ we use $\Delta _{\Omega }=0.8$ eV (blue), $\Delta
_{\Omega }=0.81$ eV (green), and $\Delta _{\Omega }=0.82$ eV (red), whereas
for $\protect\alpha _{xy}^{v}$ we use $\Delta _{\Omega }=0.6$ eV (blue), $%
\Delta _{\Omega }=0.65$ eV (green), and $\Delta _{\Omega }=0.7$ eV (red). All
other parameters are the same as in Fig.\ 1. }
\label{fig:3}
\end{figure}

\end{document}